\documentclass{elsart}
\usepackage{changebar,dsfont,float,graphicx,amsmath}
\DeclareGraphicsExtensions{.eps}
\newcommand{\reffig}[1]{Fig.~\ref{#1}}

\newif\ifpdf
\ifx\undefined\pdfoutput \pdffalse \else \pdftrue \fi
\ifpdf
  \DeclareGraphicsExtensions{.pdf}
\fi

\begin{document}
\begin{frontmatter}

\title{Line Structures in Recurrence Plots}

\author{Norbert Marwan\corauthref{cor}},
\corauth[cor]{Corresponding author.}
\author{J\"urgen Kurths}
\address{Nonlinear Dynamics Group, Institute of Physics, 
University of Potsdam, \\Potsdam 14415, Germany\\
marwan@agnld.uni-potsdam.de}

\begin{abstract}
Recurrence plots exhibit line structures which represent 
typical behaviour of the investigated system. The local slope of these
line structures is connected with a specific transformation of the time
scales of different segments of the phase-space trajectory. 
This provides us a better understanding of the structures
occuring in recurrence plots. The relationship between the
time-scales and line structures are of practical importance
in cross recurrence plots. Using this relationship within cross 
recurrence plots, the time-scales of differently sampled or
time-transformed measurements can be adjusted. An application to
geophysical measurements illustrates the capability of this
method for the adjustment of time-scales in different measurements.
\end{abstract}
\end{frontmatter}

\section{Introduction}\label{sec:intro}

In the last decade of data analysis an impressive 
increase of the application of methods based on recurrence 
plots (RP) could be observed. Introduced by Eckmann et al.~\cite{eckmann87},
RPs were firstly only a tool for the visualization
of the behaviour of phase-space trajectories. The following
development of a quantification of RPs by 
Zbilut and Webber \cite{zbilut92,webber94} and later by Marwan 
et al.~\cite{marwan2002herz}, has consolidated the
method as a tool in nonlinear data analysis. With this 
quantification the RPs have become 
more and more popular within a growing group of scientists who
use RPs and their quantification techniques for data analysis. 
Last developments
have extended the RP to a bivariate and multivariate
tool, as the cross recurrence plot (CRP) or the multivariate
recurrence plot (MRP) \cite{marwan2002pla,marwan2004nova,romano2004}. 
The main advantage of methods based on RPs is that they can also be applied to
rather short and even nonstationary data.

The initial purpose of RPs was the visual inspection of higher 
dimensional phase space trajectories. The view on RPs gives 
hints about the time evolution of these trajectories. 
The RPs exhibit characteristic large scale and small scale patterns. 
Large scale patterns can be characterized as homogeneous, periodic, 
drift and disrupted. They obtain the global behaviour of the system
(noisy, periodic, auto-correlated etc.). The quantification of RPs and CRPs 
uses the small-scale structures which are contained in these plots.
The most important ones are the diagonal and
vertical/horizontal straight lines because they reveal typical
dynamical features of the investigated system, such as range of
predictability or properties of laminarity. However, under a closer view
a large amount of bowed, continuous lines can also be found. The progression of
such a line represents a specific relationship within the data.
In this paper we present a theoretical background of this
relationship and discuss a technique to infer the 
adjustment of time-scales of two different data series.
Finally, an example from earth siences is given.

\section{Recurrence Plots}

A recurrence plot (RP) is a two-dimensional squared matrix with black 
and white dots and two time-axes, where each black dot at the
coordinates $(t_1, t_2)$ represents a 
recurrence of the system's state $\vec x(t_1)$ at time $t_2$:
\begin{equation}\label{eq:rp}
\mathbf{R}(t_1,t_2) = \Theta\left(\varepsilon - \left\|\vec x(t_1) - \vec x(t_2)\right\|\right),
\quad \vec x(t) \in \mathds{R}^m,
\end{equation}
where $m$ is the dimension of the system (degrees of freedom), 
$\varepsilon$ is a small threshold distance, \mbox{$\Vert\cdot\Vert$}
a norm and $\Theta (\cdot)$ the Heaviside function.
This definition of an RP is only one of several possibilities
(an overview of recent variations of RPs can be found in \cite{marwan2003diss}). 

Since $\mathbf{R}(t_1,t_1)=1$ by definition, the RP has a 
black main diagonal line, the {\it line of identity (LOI)},
with an angle of $\pi/4$. It has to be noted that 
a single recurrence point at $(t_1,t_2)$ in an RP does not contain any information
about the actual states at the times $t_1$ and $t_2$ in phase space. However, 
it is possible to reconstruct dynamical properties of the data from 
the totality of all recurrence points \cite{thiel2004b}.

\section{Line Structures in Recurrence Plots}

The visual inspection of RPs reveals (among other things) the following
typical small scale structures: {\it single dots}, {\it diagonal lines} as well as 
{\it vertical} and {\it horizontal lines} (the combination of vertical and horizontal 
lines plainly forms rectangular clusters of recurrence points). 

{\it Single, isolated recurrence points} can occur if states are rare, 
if they do not persist for any time, or if they
fluctuate heavily. However, they are not a clear-cut indication of chance or 
noise (for example in maps). 

A {\it diagonal line} $\mathbf{R}(t_1+\tau,t_2+\tau)=1$ (for 
$\tau=1 \dots l$, where $l$ is the length of the diagonal line
in time units) occurs
when a segment of the trajectory runs parallel to another segment,
i.\,e.~the trajectory visits the same region of the phase space at
different times. The length of this diagonal line is determined by 
the duration of such a similar local evolution of the trajectory segments. 
The direction of these diagonal structures can differ. Diagonal
lines parallel to the LOI (angle $\pi/4$) represent the parallel running of 
trajectories for the same time evolution. The diagonal structures
perpendicular to the LOI represent the parallel running with contrary
times (mirrored segments; this is often a hint of an inappropriate 
embedding if an embedding algorithm is used for the reconstruction
of the phase-space). Since the
definition of the Lyapunov exponent uses the time of the parallel 
running of trajectories, the relationship between the diagonal
lines and the Lyapunov exponent is obvious (but this relationship is
more complex than usually mentioned in literature, cf.~\cite{thiel2004a}). 

A {\it vertical (horizontal) line} $\mathbf{R}(t_1,t_2+\tau)=1$ (for 
$\tau=1 \dots v$, with $v$ the length of the vertical line in time units)
marks a time length in
which a state does not change or changes very slowly. It seems, that the state
is trapped for some time. This is a typical
behaviour of laminar states \cite{marwan2002herz}. 
%

\section{Slope of the Line Structures}

In a more general sense the line structures in an RP exhibit locally the
time relationship between the current 
trajectory segments.
A line structure in an RP of length $l$ corresponds to the closeness of 
the segment $\vec x(T_1(t))$ to another segment $\vec x(T_2(t))$, where
$T_1(t)$ and $T_2(t)$ are two local time-scales (or transformations
of an imaginary absolute time-scale $t$) which preserve that
$\vec x(T_1(t)) \approx \vec x(T_2(t))$ for some time $t=1 \dots l$. 
Under some assumptions (e.\,g.~piecewise existence of an inverse of the 
transformation $T(t)$, the two segments visit the same area in the phase 
space), a line in the RP can be simply expressed by the time-transfer
function
\begin{equation}\label{eq:los_line}
\vartheta(t) = T_2^{-1}\left(T_1(t)\right).
\end{equation} 
Especially, we find that the local slope $b(t)$ of a line in an RP represents 
the local time derivative $\partial_t$ of the inverse second time-scale $T_2^{-1}(t)$ 
applied to the first time-scale $T_1(t)$
\begin{equation}\label{eq:los}
b(t) = \partial_t T_2^{-1}\left(T_1(t)\right) = \partial_t \vartheta(t). 
\end{equation} 
This is the fundamental relation between the local slope $b(t)$ of
line structures in an RP and the time scaling of the corresponding
trajectory segments. 
From the slope $b(t)$ of a line in an RP we can infere the 
relation $\vartheta(t)$ between two segments of  $\vec x(t)$
($\vartheta(t) = \int b(t) dt$).
Note that the slope $b(t)$ depends only on
the transformation of the time-scale and is independent from the 
considered trajectory $\vec x(t)$. 

This feature is, e.\,g., used in the application of CRPs as a tool for
the adjustment of time-scales of two data series \cite{marwan2004nova,marwan2002npg} 
and will be discussed later. Next, we present 
the deforming of line structures in RPs due to different
transformations of the time-scale.

\section{Illustration Line Structures}

\begin{figure*}[htbp]
\centering{\includegraphics[width=0.47\columnwidth]{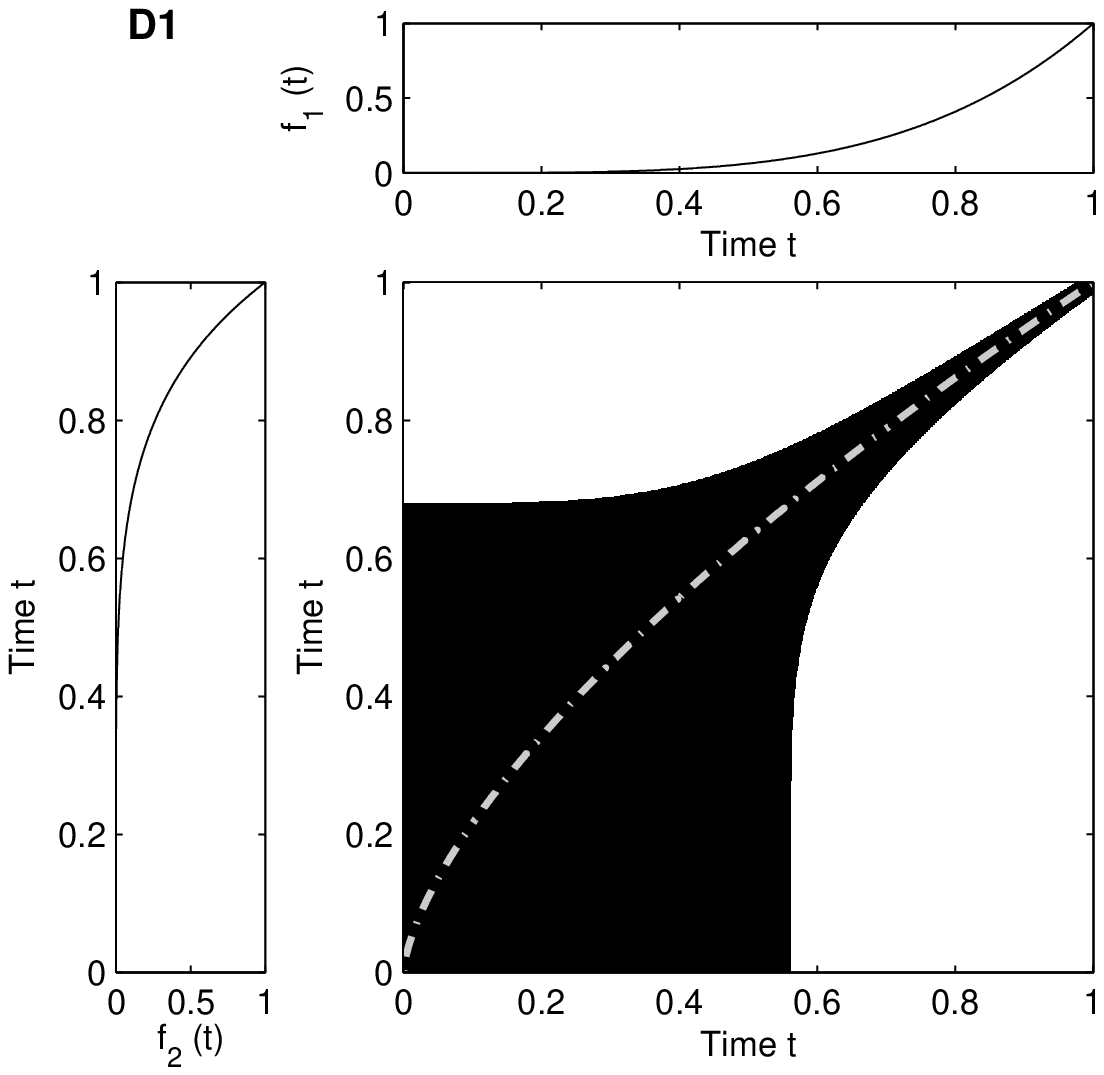}}\hspace{7mm}
\centering{\includegraphics[width=0.47\columnwidth]{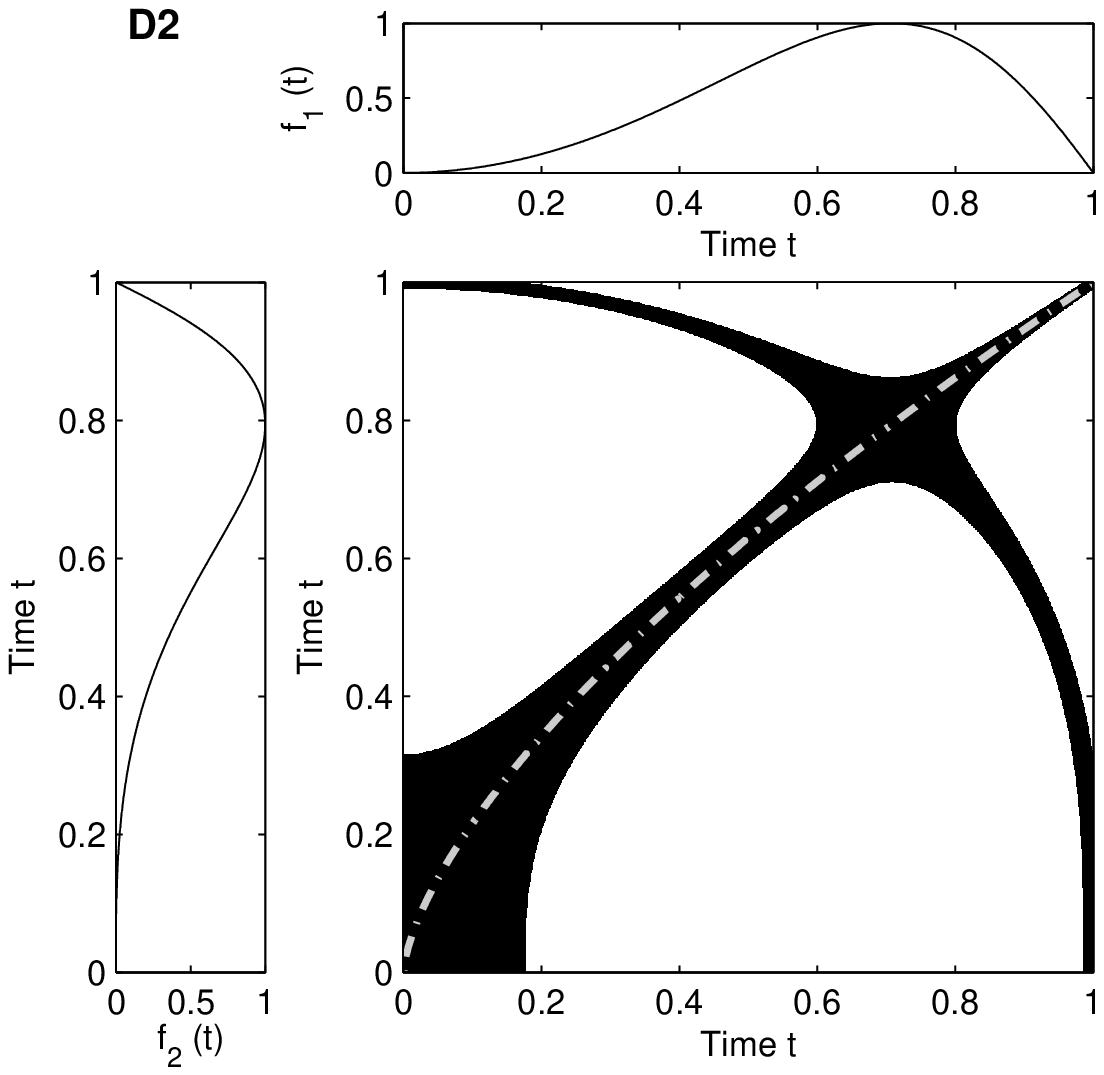}}\\ \vspace{7mm}
\centering{\includegraphics[width=0.47\columnwidth]{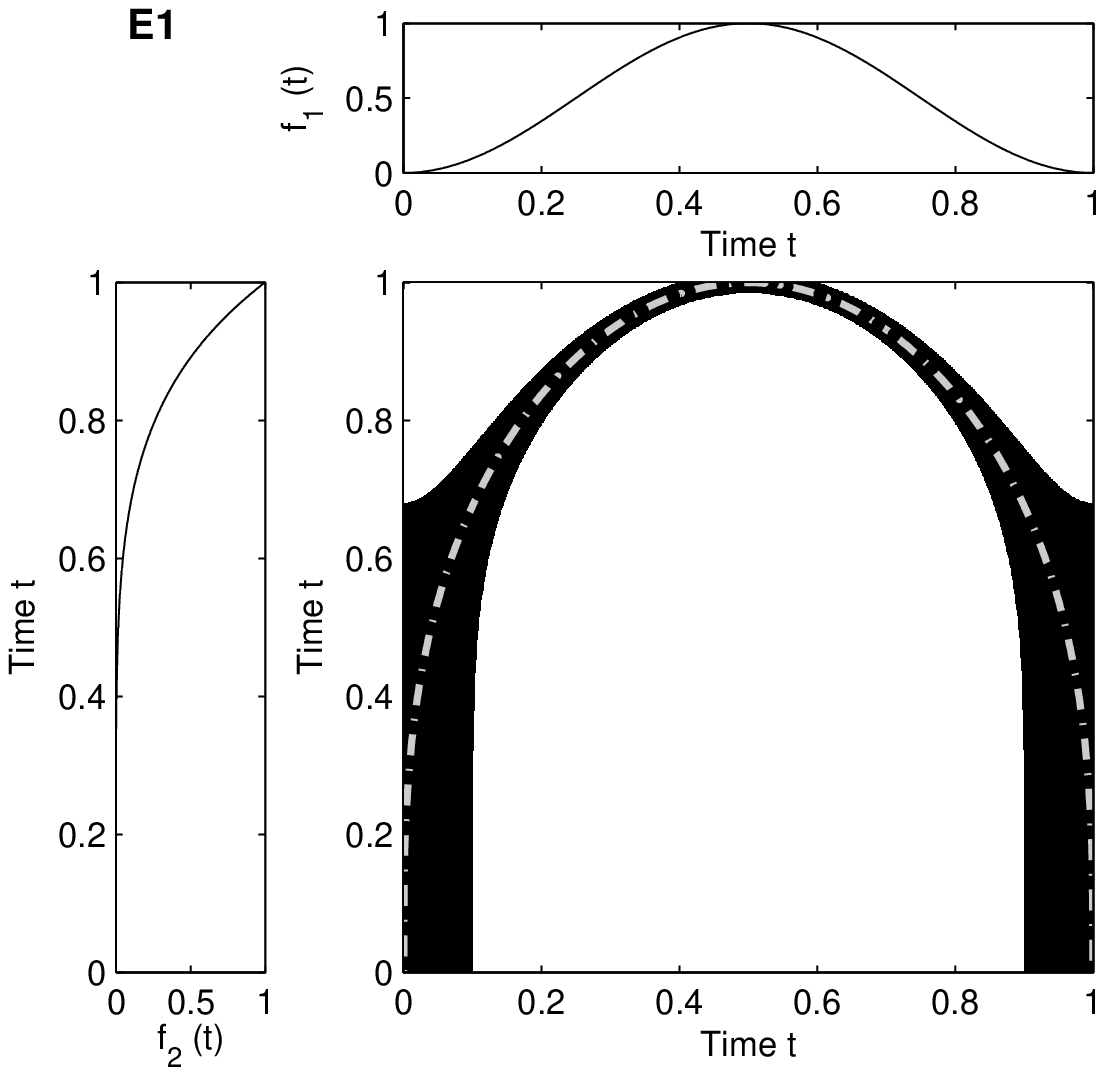}}\hspace{7mm}
\centering{\includegraphics[width=0.47\columnwidth]{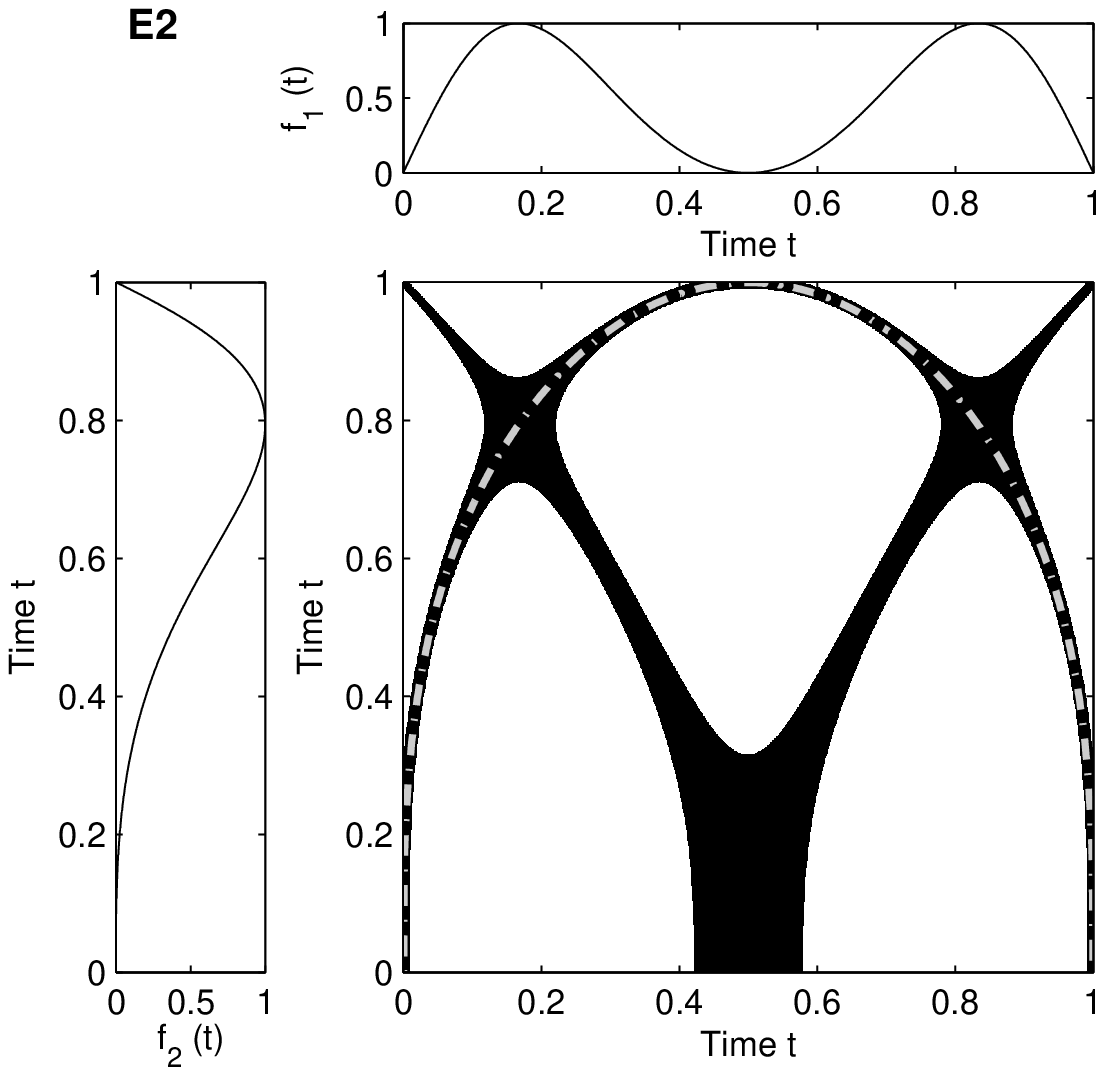}}
\caption{Details of recurrence plots for trajectories $f(t)$ 
whose sub-sections $f_1(t)$ and $f_2(t)$ undergo different 
transformations in time-scale (Tab.~\ref{tab:timetrafo}). 
Black areas correspond to times where $f_1(t) \approx f_2(t)$.
The dash-dotted lines represent the time-transfer functions $\vartheta(t)$.
Note that these are not the entire RPs, only a small detail of them
(an entire RP cannot contain only these structures -- there are
more features, like the line of identity (diagonal line from lower
left to upper right) and a more or less symmetric plot
around this line).}\label{fig:los}
\end{figure*}

For illustration we consider some examples of time transformations
for different one-dimensional trajectories $f(t)$ (i.\,e.~functions; no embedding).
We study the recurrence behaviour
between two segments $f_1$ and $f_2$ of these trajectories, where we apply 
different time transformations to these segments (Tab.~\ref{tab:timetrafo}).
In order to illustrate that the found relation (\ref{eq:los})
is independent from the underlying trajectory, we will use at first
the function $f(t) = t^2$ (Figs.~\ref{fig:los}A1, B1, C1 etc.)
and then $f(t) = \sin(\pi\,t)$ (Figs.~\ref{fig:los}A2, B2, C2 etc.) as
a trajectory.
The local representation of RPs between these segments corresponds
finally to cross recurrence plots (CRP) between two different
trajectories/functions as will be mentioned later.

\begin{table}[htbp]
\caption{Examplary time transformation functions $T_1(t)$ and $T_2(t)$,
the inverse $T_2^{-1}(t')$, their corresponding slopes $b(t)$ and 
time-transfer functions $\vartheta(t)$ for lines in RPs shown in 
Fig.~\ref{fig:los}.}\label{tab:timetrafo}
\centering \begin{tabular}{llllll}
\hline
Fig.	&$T_1(t)$	&$T_2(t)$	&$T_2^{-1}(t')$		&$b(t)$		&$\vartheta(t)$\\
\hline
\hline
A	&$t$		&$2t$		&$0.5\,t'$		&$0.5$		&$0.5\,t$\\
B	&$t$		&$5t^2$		&$\sqrt{0.2\,t'}$	&$\sqrt{\frac{0.2}{t}}$	&$\sqrt{0.2\,t}$\\
C	&$t$		&$\sqrt{1-t^2}$	&$\sqrt{1-t'^2}$	&$\frac{t}{\sqrt{1-t^2}}$	&$\sqrt{1-t^2}$\\
D	&$t^2$		&$t^3$		&$\sqrt[3]{t'}$		&$\frac{1}{3 \sqrt[3]{t^2}}$	&$\sqrt[3]{t^2}$\\
E	&$\sin(\pi\,t)$	&$t^3$		&$\sqrt[3]{t'}$		&$\frac{\pi\,\cos(\pi\,t)}{3\sqrt[3]{\sin^2(\pi\,t)}}$	&$\sqrt[3]{sin(\pi\,t)}$\\
\hline
\end{tabular}
\end{table}

\begin{figure*}[p]
\centering{\includegraphics[width=0.47\columnwidth]{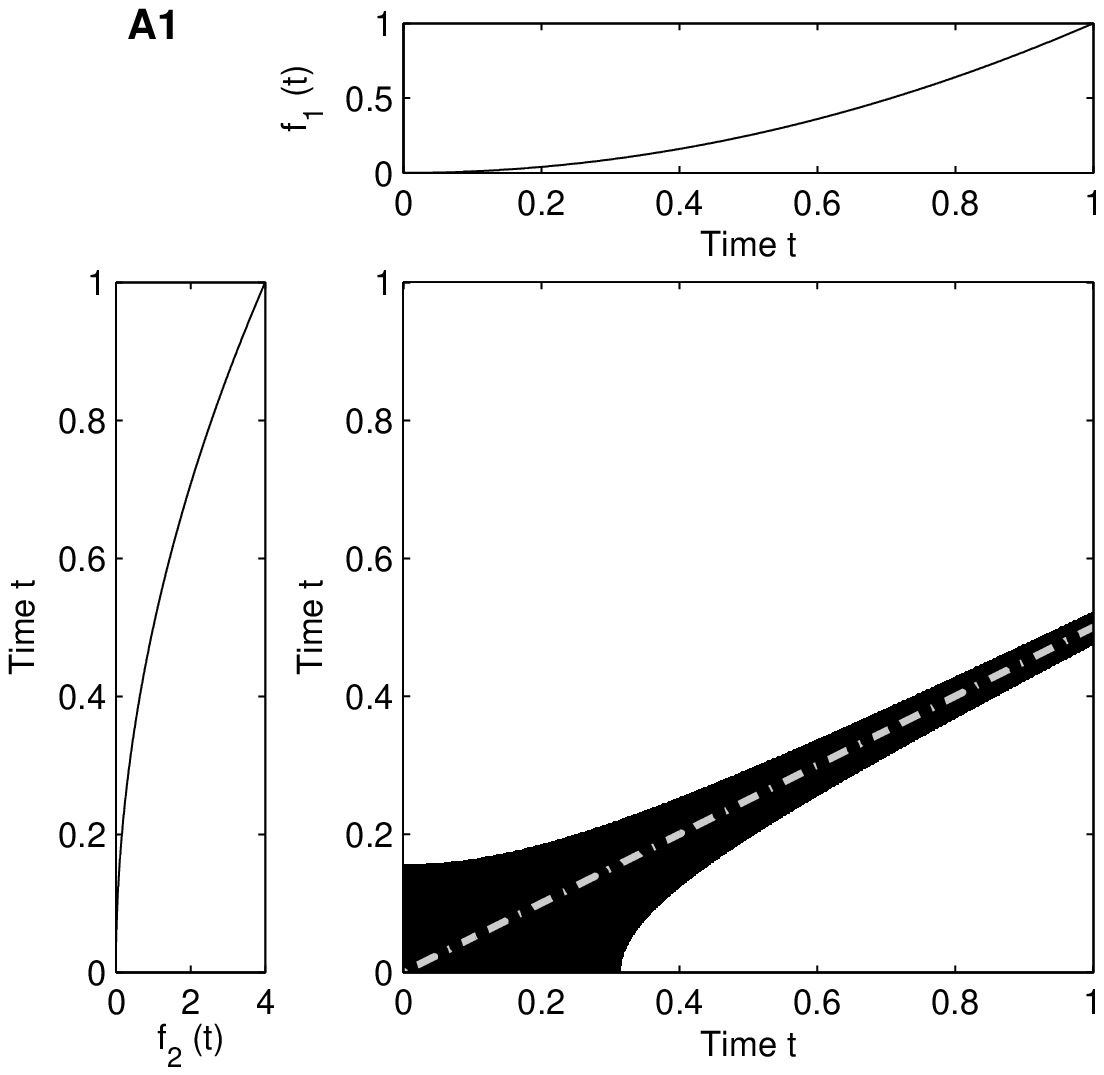}}\hspace{7mm}
\centering{\includegraphics[width=0.47\columnwidth]{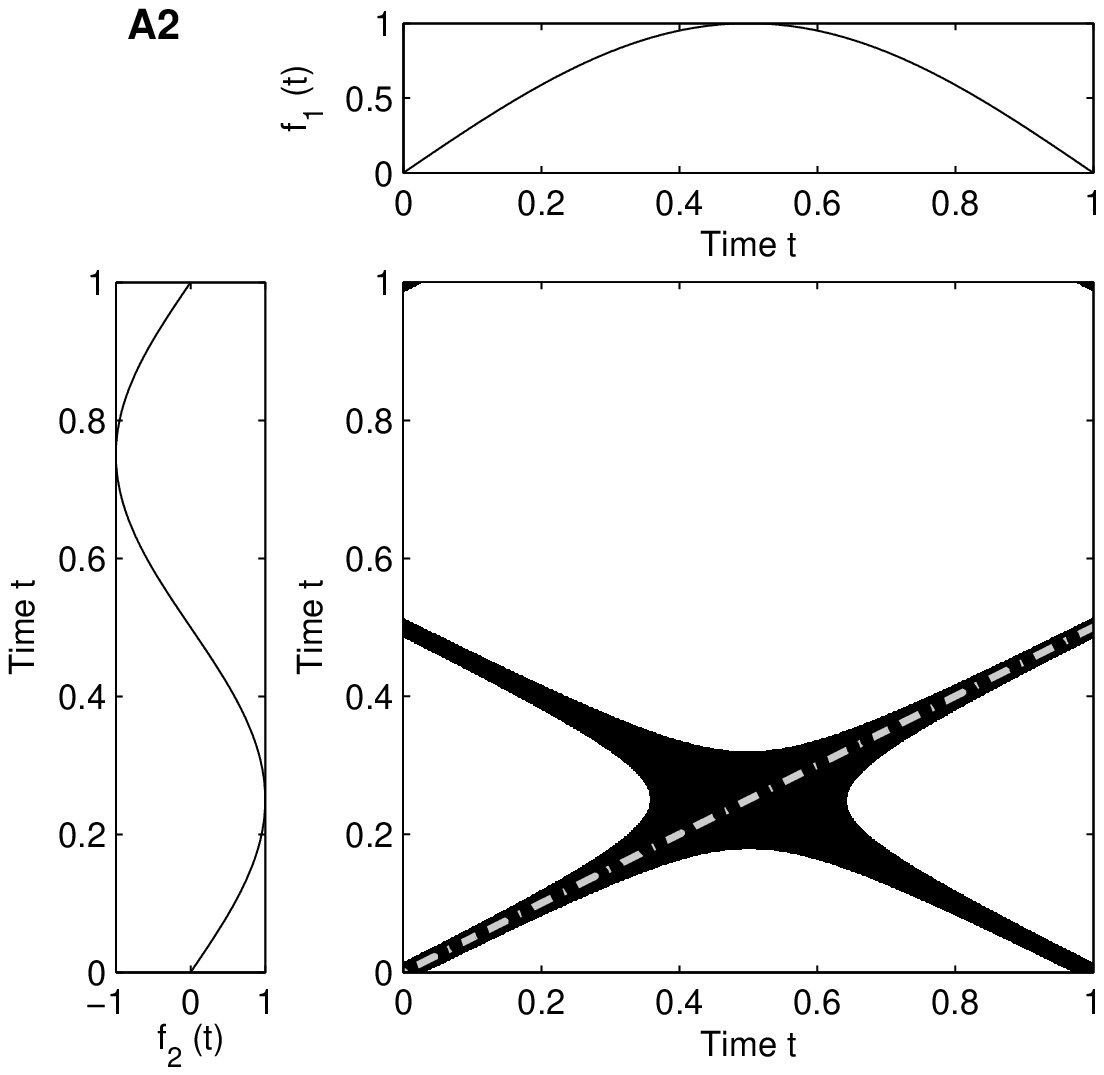}}\\ \vspace{7mm}
\centering{\includegraphics[width=0.47\columnwidth]{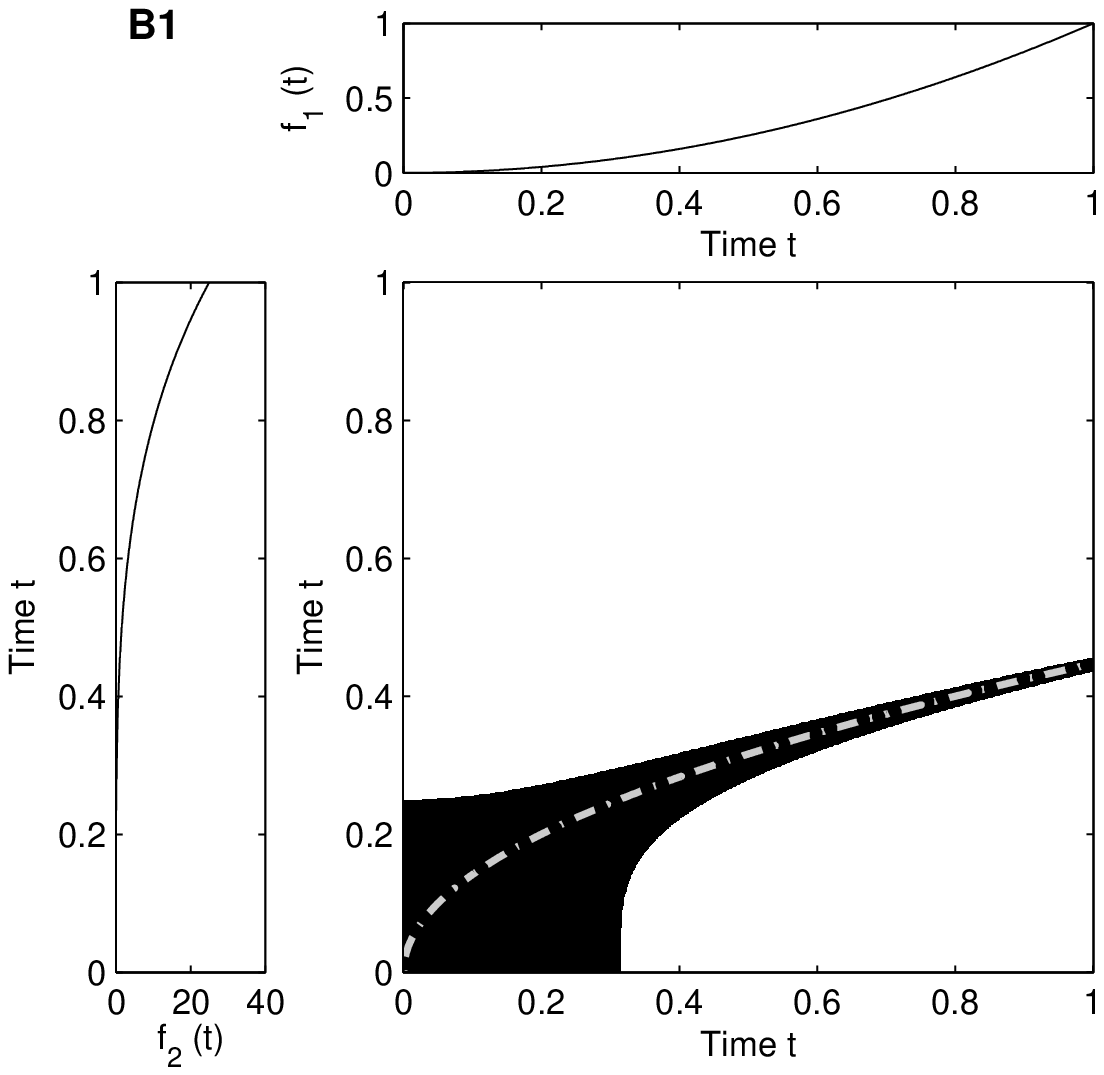}}\hspace{7mm}
\centering{\includegraphics[width=0.47\columnwidth]{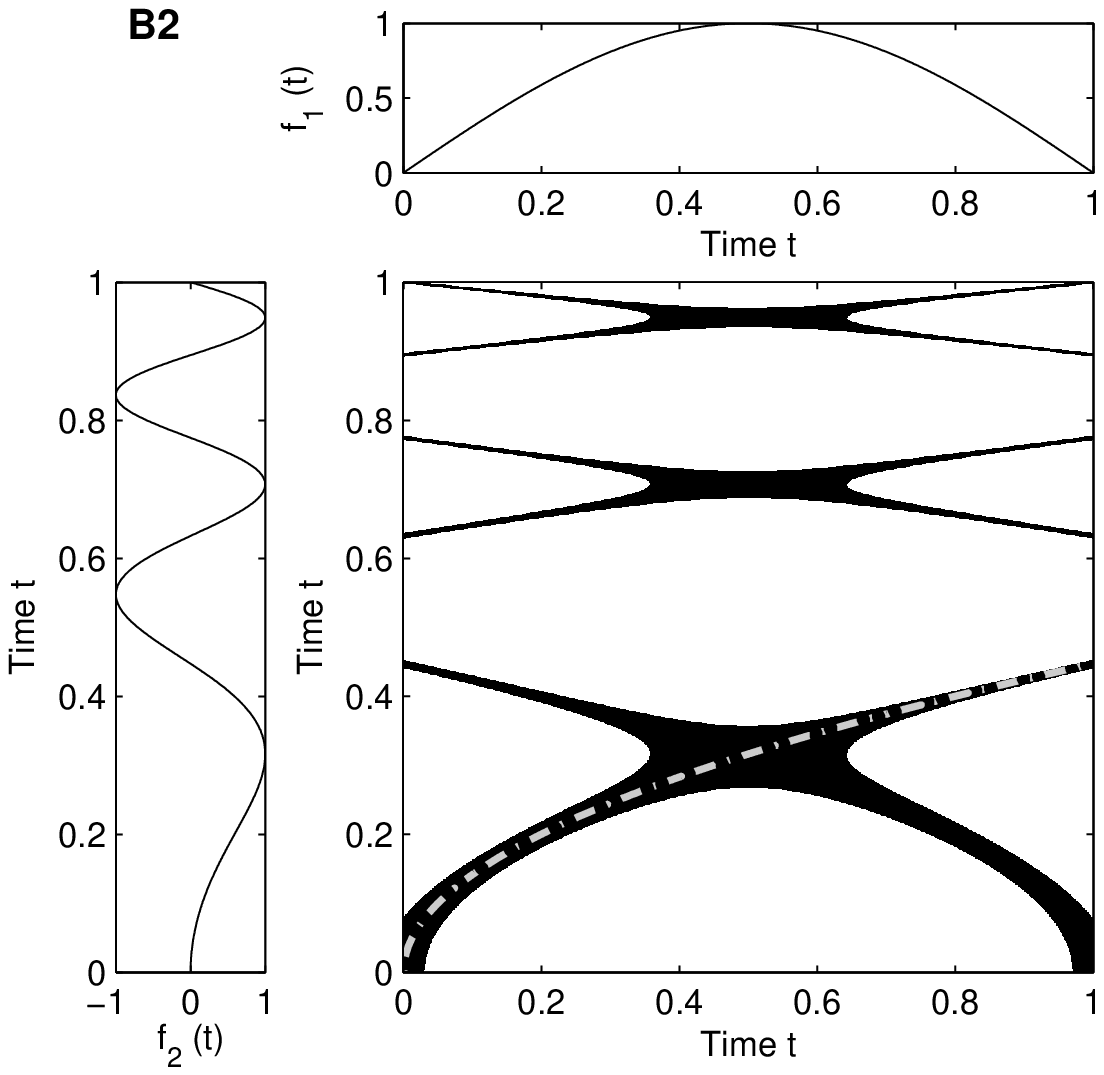}}\\ \vspace{7mm}
\centering{\includegraphics[width=0.47\columnwidth]{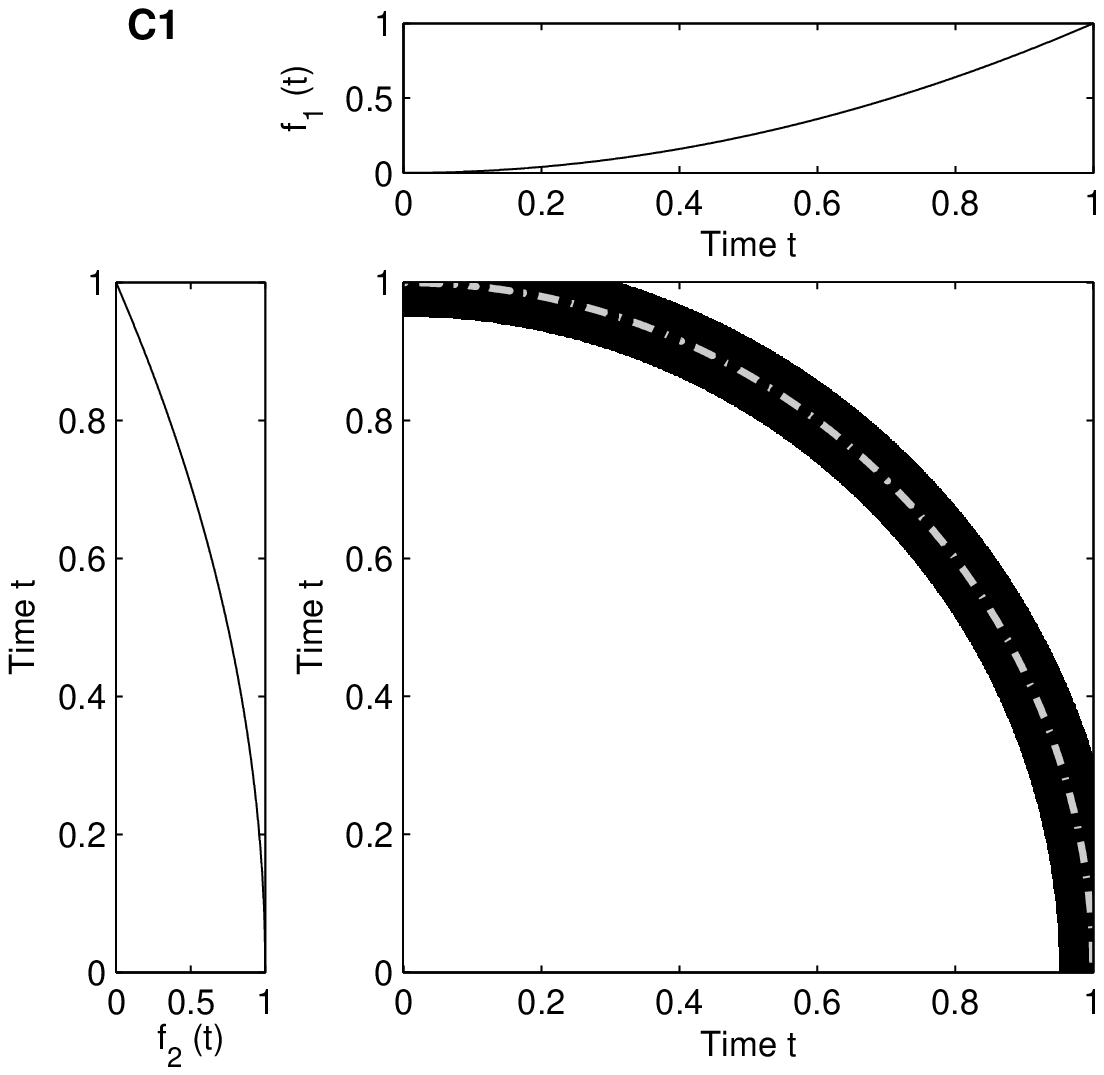}}\hspace{7mm}
\centering{\includegraphics[width=0.47\columnwidth]{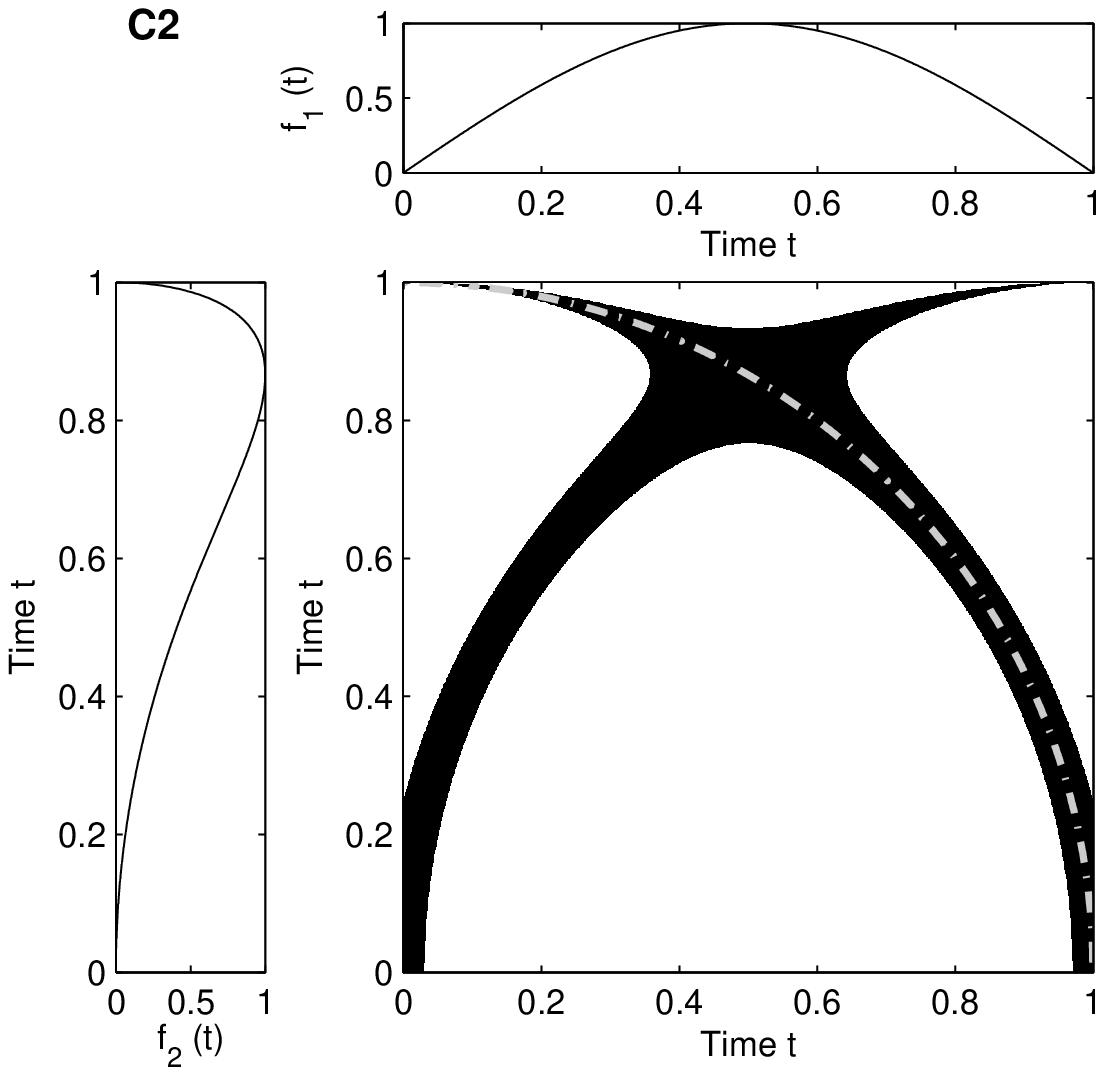}}
\end{figure*}

Assuming that the second segment of a trajectory $f_2$ is
twice as fast as the first segment $f_1$ (Figs.~\ref{fig:los}A), i.\,e.,
the time transformations are $T_1(t)=t$ and $T_2(t)=2t$, we 
get a constant slope $b=0.5$ by using Eq.~(\ref{eq:los}).
A line in an RP which corresponds to these both segements
follows $\vartheta(t) = 0.5\,t$ (Figs.~\ref{fig:los}A1, A2). 
This result corresponds with the solution we had already
discussed in \cite{marwan2002npg} using another approach.
In \cite{marwan2002npg} we considered a simple case of two harmonic 
functions $f_1(t) = \sin(T_1(t))$ and $f_2(t) = \sin(T_2(t))$
with different time transformation functions 
$T_1=\varphi\cdot t+\alpha$ and $T_2=\psi\cdot t+\beta$.
Using the inverse $T_2^{-1}=\frac{t-\beta}{\psi}$ and Eq.~(\ref{eq:los}), 
we get the local slope of lines in the RP (or CRP)
$b = \partial_t T_{2}^{-1}\left( T_{2}(t)\right)= \varphi/\psi,$
which equals the ratio between the frequencies
of the considered harmonic functions.

In the second example we will transform the time-scale of the second 
segment with the square function $T_2(t)=5t^2$. Using Eq.~(\ref{eq:los})
we get $b(t)=\sqrt{0.2/t}$ and $\vartheta(t)=\sqrt{0.2\,t}$, which 
corresponds with a bowed line in the RP (Figs.~\ref{fig:los}B1, B2). Since 
$\sin(\pi\,x)$ has some periods in the considered intervall, we get 
some more lines in the RP (Figs.~\ref{fig:los}B2). These lines 
underly the same relationship, but we have to take 
higher periodicities into account: $\vartheta(t)=\sqrt{0.2\,k\,\pi\,t}$ ($k \in \mathds{Z}$).

\begin{figure}[htbp]
\vspace*{2mm}
\centering \includegraphics[width=7cm]{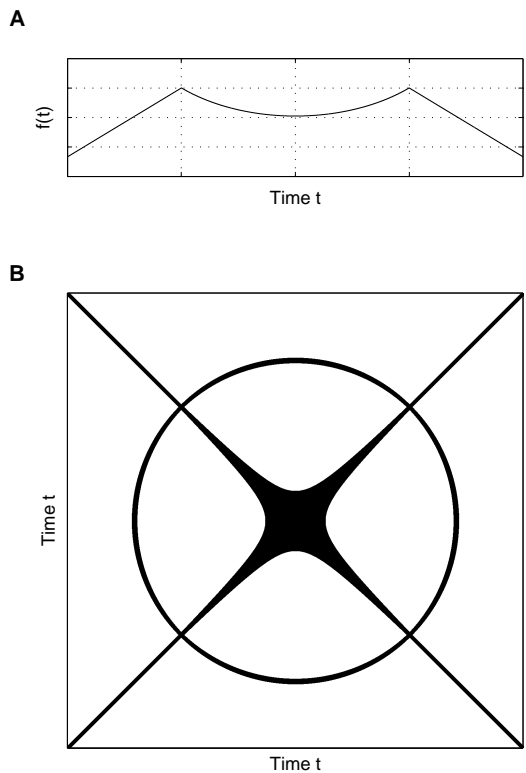}
\caption{Illustrative example of the relationship between the slope
of lines in an RP and the local derivatives of the involved trajectory 
segments. Since the local derivative of the transformation of the time-scales of the
linear and the hyperbolic sections (A) corresponds to the derivative
of a circle line, a circle occurs in the RP (B). Recurrence plot is derived from
the one-dimensional phase-space (no embedding used).}\label{fig:rp_circle}
\end{figure}

The third example refers to a hyperbolic time transformation $T_2(t)=\sqrt{1-t^2}$.
The resulting line in the RP has the slope $b(t)=t/\sqrt{1-t^2}$ and follows 
$\vartheta(t)=\sqrt{1-t^2}$, which 
corresponds with a segment of a circle (Figs.~\ref{fig:los}C1, C2).
We can use this information in order to create a full circle in an RP.
Let us consider a one-dimensional system, where the trajectory is simply the
function $f(T)=T(t)$, and with a section of a monotonical, linear 
increase $T_{lin}=t$ and another (hyperbolic) section which follows 
$T_{hyp} = -\sqrt{r^2-t^2}$. After these both sections we append 
the same but mirrored sections (\reffig{fig:rp_circle}A).
Since the inverse of the hyperbolic section is 
$T_{hyp}^{-1}=\pm \sqrt{r^2-t^2}$, the line in the corresponding
RP follows $\vartheta(t) = T_{hyp}^{-1} ( T_{lin}(t) ) = \pm \sqrt{r^2-t^2}$,
which corresponds with a circle of radius $r$ (\reffig{fig:rp_circle}B).

\begin{figure}[htbp]
\vspace*{2mm}
\centering \includegraphics[width=7cm]{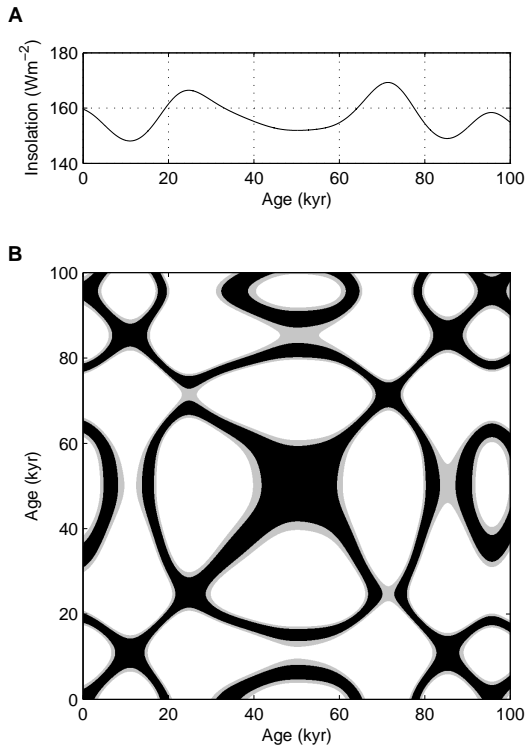}
\caption{A corresponding
structure found in experimental data: (A) the solar insolation on the latitude
44$^{\circ}$N for the last 100~kyr (data from \cite{berger91})
and its corresponding recurrence plot (B). The recurrence plot
parameters were $m=1$ and $\varepsilon=2$ (black)
and $\varepsilon=3.5$ (gray).}\label{fig:rp_inso}
\end{figure}

An examplary data series from earth science reveals that such structures
are not only restricted to artificial models. Let us consider the
January solar insolation for the last 100 kyr on the latitude 44$^{\circ}$N
(\reffig{fig:rp_inso}A). The corresponding RP 
shows a circle (\reffig{fig:rp_inso}B), similar
as in \reffig{fig:rp_circle}B. From this geometric structure we can
infer that the insolation data contains a more-or-less
symmetric sequence and that subsequent sequences are equal after a
suitable time transformation which follows the relation 
$T_2^{-1}(T_1) = \sqrt{r^2-t^2}$. For instance, the subsequent 
sequences could be a linear increasing and a hyperbolic decreasing
followed by a reverse of this sequence, a hyperbolic increasing and 
a linear decreasing part. Such bowed line structures can also be found 
in RPs applied to data from biology, ecology, economy.
These deformations can obtain hints about the change of frequencies
during the evolution of a process and may be of major interest especially
in the analysis of sound data (an example of an RP of speech
data containing pronounced bowed lines can be found in \cite{hegger2000a}).

Whereas in the examples above only the second section of the trajectory
undergoes a time transformation, in the last two examples 
(Figs.~\ref{fig:los}D and E) the time-scale of the first section is also
transformed. Nevertheless, the time-transfer function can
be again determined with Eq.~(\ref{eq:los_line}) too. 

From these examples we can conclude that the line in a recurrence
plot follows Eq.~(\ref{eq:los_line}) and depends only on the 
transformations of the time-scale.

\section{Cross Recurrence Plots}

The relationship between the local slope of line structures in RPs
and the corresponding different segments of the {\it same}
phase-space trajectory holds also for the structures in
CRPs,
\begin{equation}\label{eq:crp_ex}
\mathbf{CR}(t_1, t_2)=\Theta\bigl(\varepsilon-\left\| \vec x(t_1)-\vec y(t_2)\right\|\big) .
\end{equation}
which are based on {\it two different} phase-space trajectories $\vec x(t_1)$ and $\vec y(t_2)$.
This relationship is more important for the {\it line
of identity (LOI)} which then becomes a {\it line of 
synchronization (LOS)} in a CRP \cite{marwan2004nova,marwan2002npg}.

We start with two identical trajectories, i.\,e.~the CRP is the same as the
RP of one trajectory and contains an LOI. If we now slightly modify
the amplitudes of the second trajectory, the LOI
will become somewhat disrupted. This offers a new approach to use CRPs
as a tool to assess the similarity of two systems 
\cite{marwan2002pla}. However, if we do not modify the amplitudes but
stretch or compress the second trajectory slightly, the LOI will
remain continuous but not as a straight line with an angle of $\pi/4$.
The line of identity (LOI) now becomes the 
{\it line of synchronization (LOS)} and may eventually not have
the angle $\pi/4$. This line can be rather bowed. 
Finally, a time shift between the trajectories causes a 
dislocation of the LOS, hence, the LOS may lie rather far from the main 
diagonal of the CRP.

Now we deal with a situation which is typical in earth sciences and assume
that two trajectories represents the same process but contain some transformations
in their time-scales. The LOS in the CRP between the two trajectories 
can be described with the found relation 
(\ref{eq:los_line}). The function $\vartheta(t)$ is the transfer or rescaling 
function which allows to readjust the time-scale of the 
second trajectory to that of the first one in a non-parametrical way. 
This method is useful for all tasks where two time-series have to be adjusted
to the same scale, as in dendrochronology or sedimentology \cite{marwan2004nova}.

\begin{figure}[htbp]
\vspace*{2mm}
\centering \includegraphics[width=\columnwidth]{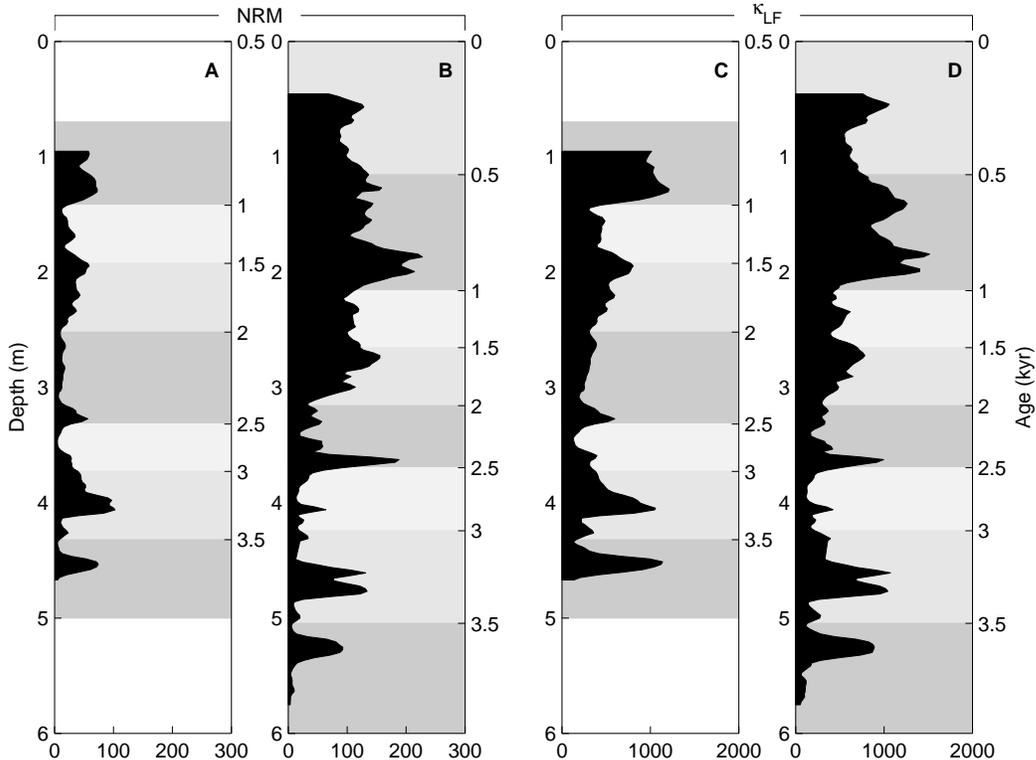}
\caption{Rock-magnetic measurements of lake sediments with different
time-scales. Corresponding sections are marked with different gray values.}\label{fig:sample_raw}
\end{figure}

Next, we apply this technique in order to re-adjust two geological
profiles (sediment cores) from the Italian lake {\it Lago di Mezzano} \cite{brandt99}.
The profiles cover approximately the same geological processes but have 
different time-scales due to variations in the sedimentation rates.
The first profile (LMZC) has a length of about 5~m and the second
one (LMZG) of about 3.5~m (\reffig{fig:sample_raw}). From both profiles
a huge number of geophysical and chemical parameters were measured. 
Here we focus on the rock-magnetic measurements of the normalized
remanent magnetization intensity (NRM) and the susceptibility $\kappa$.

\begin{figure}[htbp]
\vspace*{2mm}
\centering \includegraphics[width=10cm]{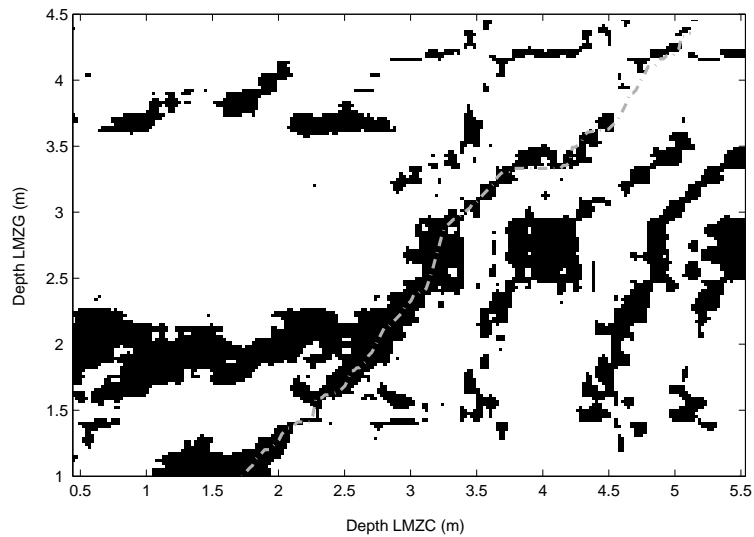}
\caption{Cross recurrence plot between rock-magnetic data shown in Fig.~\ref{fig:sample_raw}.
The dash-dotted line is the resolved LOS which can be used for re-adjustment the time-scales
of both data sets.}\label{fig:sample_rp}
\end{figure}

\begin{figure}[htbp]
\vspace*{2mm}
\centering \includegraphics[width=\columnwidth]{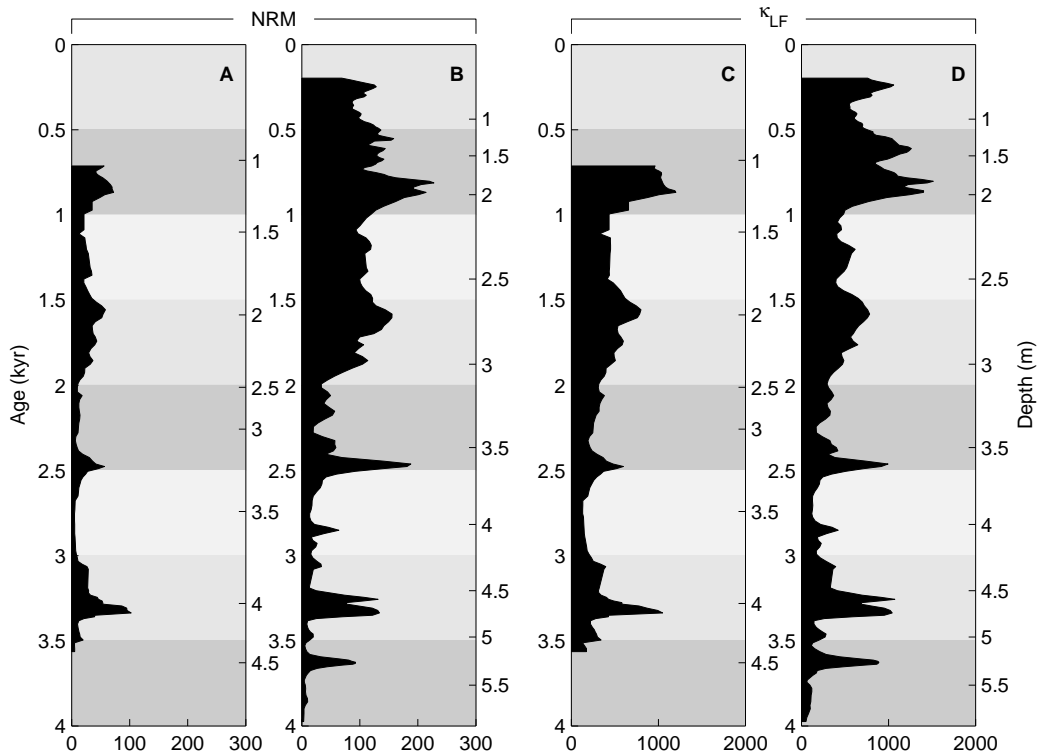}
\caption{Geological profiles after re-adjustment using the LOS which was found with
the CRP shown in Fig.~\ref{fig:sample_rp}. Corresponding sections are marked with 
different gray values.}\label{fig:sample_adjust}
\end{figure}

We use the time-series NRM and $\kappa$ as components for the
phase-space vector, resulting in a two-dimensional system.
However, we apply an additional embedding using the time-delay
method \cite{theiler92} (we do not ask about the physical meaning here). 
A rather small embedding decreases the amount
of line structures representing the progress with negative time \cite{marwan2003diss}.
Using embedding parameters dimension $m=3$ and delay $\tau=5$ (empirically found for these
time-series), the final dimension of the reconstructed system is six.
The corresponding CRP reveals a partly disrupted, swollen and bowed LOS (\reffig{fig:sample_rp}).
This LOS can be automatically resolved, e.\,g.~by using the LOS-tracking algorithm
as described in \cite{marwan2002npg}. The application of this LOS
as the time-transfer function to the profile LMZG re-adjusts its time-series
to the same time-scale as LMZC (\reffig{fig:sample_adjust}).
This method offers a helpful tool for an automatic adjustment
of different geological profiles, which offers advantages compared to the rather subjective
method of "wiggle matching" (adjustment by harmonizing maxima and minima by eye) used so far.

\section{Conclusion}
Line structures in recurrence plots (RPs) and cross recurrence plots (CRPs)
contain information about epochs of a similar evolution
of segments of phase-space trajectories. Moreover the local slope of such
line structures is directly related with the difference in
the velocity the system changes at different times.
We have demonstrated that the knowlege about this relationship allows a better understandig
of even bowed structures occuring in RPs. This relationship can be used to 
analyse changes in the time domain of data series (e.\,g.~frequencies), as it is
of major interest, e.\,g., in the analysing of speech data.
We have used this feature 
in a CRP based method for the adjustment of time-scales between different 
time-series. The potential of this technique is finally shown for experimental
data from geology.

\section{Acknowledgements}
This work was partly funded or supported by the Special Research Programmes SPP1097 and SPP1114
of the German Science Foundation (DFG) as well as by the project AO-99-030
of European Space Agency.
We gratefully acknowledge N.~Nowaczyk and U.~Frank (GeoForschungsZentrum Potsdam) for the helpful
discussions and for providing the geophysical data. The recurrence plots and
cross recurrence plots were created by using the CRP toolbox 
for Matlab (http://tocsy.agnld.uni-potsdam.de).

\bibliographystyle{elsart-num}
\bibliography{../mybibs,../rp}

\end{document}